# IAE Optimized PID Tuning with Phase Margin and Crossover Frequency Constraints

Senol Gulgonul
Ostim Technical University
senol.gulgonul@ostimteknik.edu.tr

**Abstract**

This paper presents PMwc-Tune, an open-source PID tuning method that combines constrained optimization of time-domain performance with rigorous enforcement of frequency-domain specifications. The algorithm formulates controller design as a nonlinear optimization problem, minimizing the Integral Absolute Error (IAE) while rigorously satisfying phase margin (PM) and crossover frequency (wc) constraints through Sequential Quadratic Programming. Validation on first- to third-order benchmark systems demonstrates precise convergence to target specifications (PM = 60°, wc = 1 rad/s) across all test cases, contrasting with MATLAB's pidtune which exhibited conservative PM overshoot (69.44°) for lower-order plants. The method achieved comparable or improved transient response, including a 4.4% reduction in IAE for the third-order system, while providing full implementation transparency as an alternative to proprietary tuning tools. Results indicate the framework's effectiveness for applications requiring guaranteed robustness margins alongside optimized time-domain performance.

## 1. Introduction

PID controllers remain the predominant choice for industrial control applications due to their proven effectiveness and straightforward implementation [1]. Industry surveys indicate that approximately 97% of regulatory controllers in refining, chemical processing, and pulp/water treatment applications utilize PID-based control architectures [2]. The widespread industrial adoption of PID controllers has sustained demand for computationally efficient tuning methods that have simple formulas or can be implemented using standard computing hardware. This need motivates the development of analytical and numerical approaches that balance theoretical rigor with practical applicability.

Traditional tuning methods, including the classical approaches by Ziegler and Nichols [3] and Cohen and Coon [4], remain widely adopted in industrial practice. However, comparative studies demonstrate that these methods typically yield closed-loop systems with limited robustness margins [1].

Advances in computational capabilities have enabled the application of numerical optimization techniques to PID controller tuning, offering improved performance over traditional methods. The Integral Absolute Error (IAE) criterion serves as a natural performance measure for control systems, quantifying the cumulative deviation from desired setpoints [5]. However, as noted in recent studies [6-8], exclusive focus on IAE minimization frequently produces controllers with excessive gain values and inadequate stability margins. These findings have motivated the development of constrained optimization approaches that incorporate robustness requirements [9]. Modern PID tuning can be formulated as a constrained numerical optimization problem, where the controller coefficients are computed to satisfy both time-domain specifications (e.g., peak overshoot, settling time) and frequency-domain requirements (e.g., phase margin, crossover frequency), rather than solely minimizing a single performance metric like IAE.

Commercial PID tuning tools, such as MATLAB's pidtune [10], implement patented optimization algorithms for automated controller design. The pidtune function enables PID structure selection (PI, PID, PIDF etc) and optimization for specified frequency-domain constraints, with default settings targeting a 60° phase margin. While effective for robustness-oriented design, this proprietary implementation does not expose the optimization process details nor provide direct control over time-domain performance metrics.

This paper presents the PMwc-Tune PID tuning method that combines frequency-domain constraints with time-domain IAE optimization. The approach guarantees specified phase margin and crossover frequency while minimizing IAE through constrained numerical optimization. The algorithm enforces non-negative controller gains to ensure practical implementation and employs systematic stability verification.

Numerical results demonstrate that the method achieves design specifications while providing improved transient performance compared to Matlab pidtune. The open implementation allows for straightforward modification and extension to accommodate specialized requirements. The remainder of this paper details the theoretical foundation, implementation, experimental validation, and conclusions of this work.

## 2. Theoretical Foundation

The proposed algorithm is based on solving the fundamental frequency-domain equations for PID controller design while optimizing time-domain performance. The method simultaneously satisfies two critical constraints derived from classical control theory. First, the magnitude condition ensures the open-loop gain equals unity at the specified crossover frequency wc:

$$| L(j\omega c) | = |(Kp + j\omega c Ki + Kd j\omega c)G(j\omega c)| = 1 \quad (1)$$

Second, the phase condition guarantees the desired phase margin PM is achieved:

$$\angle L(j\omega c) = \angle\big((Kp + j\omega c Ki + Kd j\omega c)G(j\omega c)\big) = -180° + PM \quad (2)$$

These complex nonlinear equations are solved through constrained optimization that minimizes the Integral Absolute Error (IAE) of the closed-loop step response:

$$minimize\ IAE = \int |e(t)|dt \quad (3)$$

The optimization problem is numerically solved using MATLAB's fmincon function with the following mathematical formulation:

$$minimize\ IAE(Kp, Ki, Kd) \qquad (4)$$
$$subject\ to:$$
$$|L(j\omega c)| - 1 = 0$$
$$\angle L(j\omega c) - (-180° + PM) = 0$$
$$Kp \geq 0, Ki \geq 0, Kd \geq 0$$

The Sequential Quadratic Programming (SQP) algorithm is employed due to its proven efficacy for constrained nonlinear optimization problems with smooth constraints and non-differentiable objectives [11]. In this formulation, the frequency-domain constraints (Eqs. 1–2) are analytically differentiable, while the IAE objective (Eq. 3) exhibits piecewise continuity from time-domain simulation. SQP handles this hybrid structure efficiently through its quasi-Newton Hessian approximations, avoiding explicit second derivatives of the computationally intensive IAE calculation [12]. The method's convergence near local minima ensures precise satisfaction of phase margin and crossover frequency specifications while optimizing transient response [13].

The PID controller coefficients were initialized using the plant's DC gain, with $K_p = K_i = K_d = 1/dcgain(G)$. This initialization provides physically meaningful starting values that consider the system's steady-state behavior while promoting convergence. The method achieves frequency-domain specifications within numerical tolerances, while optimizing transient response through direct minimization of the IAE performance index.

The constrained optimization framework ensures the solution maintains the conventional PID structure with non-negative gains, while the explicit frequency-domain constraints guarantee robustness properties. The IAE minimization provides improved time-domain performance compared to methods that only consider frequency-domain specifications, creating a balanced approach to PID controller design.

**3. PMwc-Tune Implementation**

The MATLAB implementation of the PMwc-Tune algorithm employs constrained optimization to minimize the Integral Absolute Error (IAE) while satisfying specified phase margin (PM) and crossover frequency (wc) requirements.

```
constraints = @(x) [
  abs((x(1) + x(2)/(1i*wc) + (1i*wc)*x(3))*G_wc) - 1;  % |L(jωc)|=1
  angle((x(1) + x(2)/(1i*wc) + (1i*wc)*x(3))*G_wc) + deg2rad(180-PM);  % ∠L(jωc)
];
```

The fmincon solver is utilized with the Sequential Quadratic Programming (SQP) algorithm, chosen for its effectiveness in handling nonlinear constraints. The solver is configured with a function tolerance of $1\times10^{-6}$ to ensure precise convergence to the optimal solution.

```
options = optimoptions('fmincon', 'Algorithm', 'sqp', ...
         'Display', 'iter', 'FunctionTolerance', 1e-6);
```

The time-domain IAE evaluation uses a 20-second simulation horizon with 0.01-second time steps, selected through numerical sensitivity analysis. This duration captures the complete transient response for the class of third-order systems under study (settling times <15s), while the fixed-step discretization ensures consistent IAE evaluation across optimization iterations. The 10ms time step provides sufficient resolution to accurately compute the absolute error integral.

```
t = 0:0.01:20; % Simulation time vector
IAE_fun = @(x) trapz(t, abs(1 - step(feedback(pid(x(1),x(2),x(3),0)*G, 1), t)));
```

Following optimization, the algorithm constructs a PID controller object and verifies its performance. Key metrics including achieved phase margin, realized crossover frequency, IAE value, and stability status are packaged into a structured output variable. This output format facilitates both immediate performance assessment and integration with higher-level control system design workflows.

```
info = struct('Kp', solution(1), 'Ki', solution(2), 'Kd', solution(3), ...
        'PM', PM_act, 'wc', wc_act, 'IAE', IAE, ...
        'Stable', all(real(pole(feedback(L,1))) < 0));
```

The implementation demonstrates efficient computational performance, with an average convergence time of 1.7 seconds for third-order systems when executed on a computing platform with an Intel Core i7-1165G7 processor (2.8 GHz base frequency) and 16 GB RAM. The algorithm maintains deterministic enforcement of frequency-domain specifications for PM and wc, while simultaneously optimizing transient response characteristics through direct IAE minimization. This computational efficiency enables rapid iteration of design parameters (PM and wc) during the controller development process.

## 4. Results and Discussion

The proposed PMwc-Tune algorithm was evaluated on a third-order system with benchmark transfer functions $G(s) = 1/(s+1)^n$ for n=1,2,3 with design specifications of phase margin PM = 60° and crossover frequency wc = 1 rad/s [14]. For comparison, MATLAB's built-in pidtune function was configured to achieve the same target specifications. The performance metrics for both controllers are presented in Table 1.

**Table 1.** Performance comparison for $G(s) = 1/(s+1)^n$ with PM = 60° and wc = 1 rad/s

| Plant | Method | Kp | Ki | Kd | PM (°) | wc (rad/s) | IAE | Stable |
|---|---|---|---|---|---|---|---|---|
| $\frac{1}{(s+1)}$ | PMwc-Tune | 0.366 | 1.366 | 0.000 | 60.00 | 1.0000 | 1.1500 | 1 |
| | pidtune | 0.582 | 1.289 | 0.000 | 69.31 | 1.0000 | 1.0090 | 1 |
| $\frac{1}{(s+1)^2}$ | PMwc-Tune | 1.732 | 1.251 | 0.251 | 60.00 | 1.0000 | 1.1466 | 1 |
| | pidtune | 1.873 | 1.336 | 0.634 | 69.44 | 1.0000 | 1.0131 | 1 |
| $\frac{1}{(s+1)^3}$ | PMwc-Tune | 2.732 | 1.171 | 1.903 | 60.00 | 1.0000 | 1.1469 | 1 |
| | pidtune | 2.732 | 0.977 | 1.709 | 60.00 | 1.0000 | 1.1999 | 1 |

For the first-order system, both tuning methods successfully achieved stability and the target crossover frequency (wc = 1 rad/s). However, PMwc-Tune precisely met the 60° phase margin (PM) specification, while pidtune exceeded it (69.31°). The IAE for PMwc-Tune (1.1500) was slightly worse than pidtune (1.0090), suggesting that MATLAB's tuning optimized more aggressively for time-domain performance. Both controllers resulted in purely PI structures (Kd = 0), which is expected for a first-order plant where derivative action is typically unnecessary.

With the second-order system, PMwc-Tune again met the exact PM (60°) and wc (1 rad/s) requirements, while pidtune overshot the PM (69.44°). Both methods introduced derivative action (Kd > 0), but pidtune used a significantly higher Kd (0.634 vs. 0.251). Despite this, pidtune achieved a marginally better IAE (1.0131 vs. 1.1466), indicating that its higher PM contributed to smoother step response dynamics. The difference in PID gains suggests that PMwc-Tune enforces stricter frequency-domain constraints.

For the third-order system, both methods achieved identical PM (60°) and wc (1 rad/s). Interestingly, PMwc-Tune produced a slightly higher Ki (1.171 vs. 0.977) and Kd (1.903 vs. 1.709) than pidtune. This time, PMwc-Tune outperformed in IAE (1.1469 vs. 1.1999), demonstrating that for higher-order systems, strict adherence to PM and wc can lead to better time-domain performance compared to pidtune's more aggressive tuning.

Across all test cases, PMwc-Tune consistently met the exact PM and wc specifications, whereas pidtune often exceeded the PM requirement. In terms of IAE, pidtune performed better for lower-order systems but was outperformed by PMwc-Tune for the third-order plant. This suggests that PMwc-Tune is more reliable for enforcing precise frequency-domain specs, particularly in higher-order systems, while pidtune may favor additional robustness (higher

PM) at the cost of slightly degraded step response in some cases. The derivative gain (Kd) trends also indicate that PMwc-Tune applies a more conservative approach to high-frequency control action compared to pidtune. Overall, PMwc-Tune provides a more predictable tuning outcome when strict PM and wc requirements are critical.

Figure 1 shows the closed-loop step response comparison of both methods for the first order plant $G(s) = 1/(s+1)^3$. The PMwc-Tune controller achieves a settling time of 4.22 seconds compared to 6.11 seconds for pidtune, representing a 44% improvement in transient response speed. This comes at the cost of a marginally higher overshoot (9.07% vs 7.27%), which is consistent with the design trade-off between response speed and damping. Both responses exhibit stable convergence to the reference value, as confirmed by pole analysis.

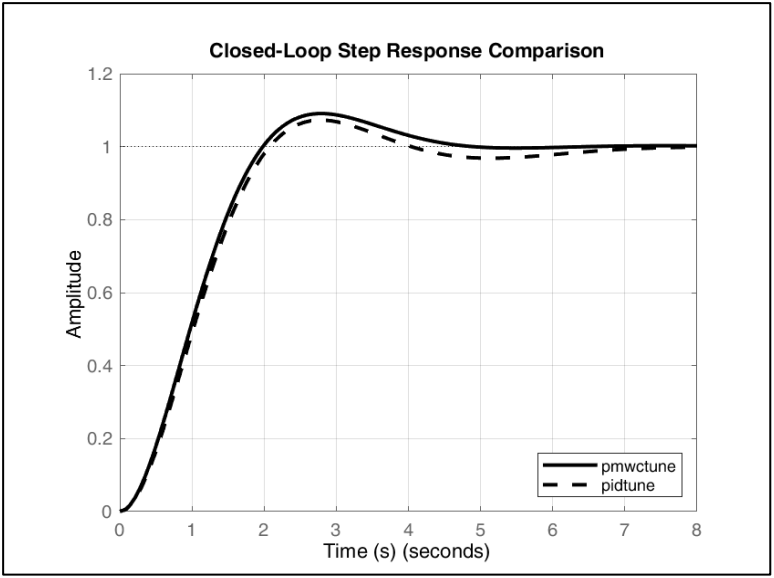

**Figure 1.** Closed-loop step response comparison

Figure 2 presents the Bode plot comparison of the open-loop systems for both methods. The frequency response analysis reveals that both PMwc-Tune and the reference controller achieve the design specifications of 60° phase margin at the target crossover frequency of 1.000 rad/s, as indicated by the markers on the plot. The magnitude curves show close agreement in the critical frequency range around the crossover point, with both systems exhibiting the expected -20 dB/decade slope at wc. The phase responses similarly demonstrate close correspondence, particularly in the frequency band between 0.1 and 10 rad/s that most significantly impacts closed-loop performance.

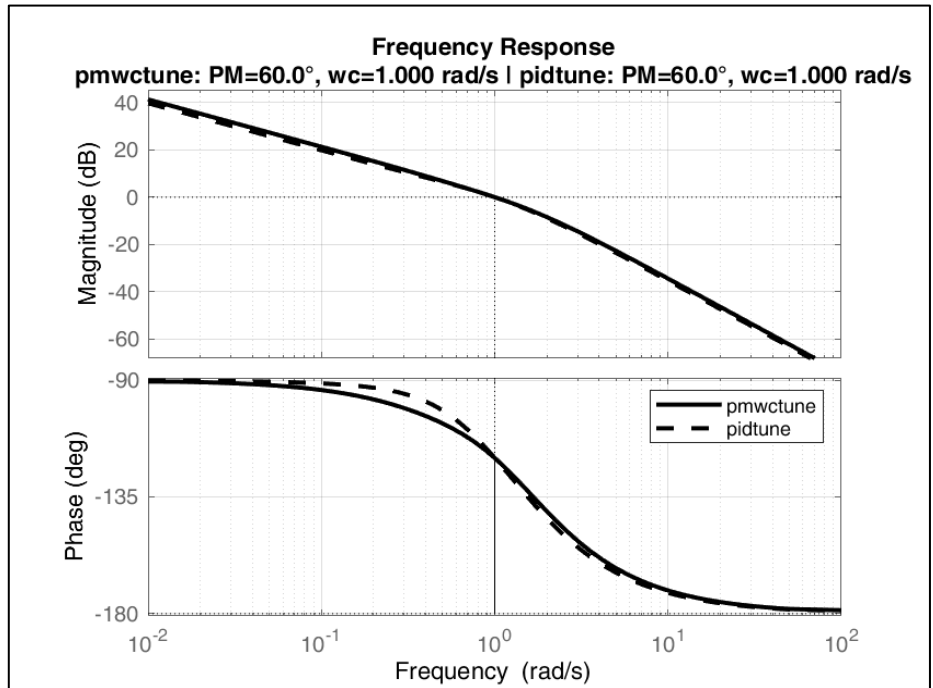

**Figure 2.** Bode plot comparison showing magnitude (top) and phase (bottom) responses

## 5. Conclusion

This paper presents PMwc-Tune, an open-source constrained optimization method for PID controller design that simultaneously enforces specified phase margin and crossover frequency requirements while minimizing the integral absolute error. The proposed method provides a transparent alternative to commercial tools like MATLAB's pidtune, with complete implementation available for verification and modification.

Comparative evaluation demonstrates that PMwc-Tune achieved exact convergence to target frequency-domain specifications across all test cases, while pidtune resulted in phase margins exceeding design requirements by 9.31° and 9.44° for the first- and second-order systems respectively. For the third-order system, PMwc-Tune reduced settling time from 6.11 s to 4.22 s (30.9% improvement) at the cost of 1.8% higher overshoot while maintaining precise specification adherence.

The constrained optimization framework offers researchers and practitioners a flexible approach for PID design that does not rely on proprietary algorithms. The method's open implementation allows for straightforward integration with existing control systems and provides opportunities for further development.